# SegmentAnyMuscle: A universal muscle segmentation model across different locations in MRI

Roy Colglazier, MD[*,1]; Jisoo Lee, MS[*,2]; Haoyu Dong, MS[*,3]; Hanxue Gu, BS[3]; Yaqian Chen, MS[3]; Joseph Cao, MD[1]; Zafer Yildiz, BS[2]; Zhonghao Liu, BS[3]; Nicholas Konz, BS[3]; Jichen Yang, PhD[3]; Jikai Zhang, MB[3]; Yuwen Chen, MS[3]; Lin Li, MS[2]; Adrian Camarena, MD[5]; Maciej A. Mazurowski, PhD[1,2,3,6]

[1]Department of Radiology, Duke University, NC, 27703, USA; [2]Department of Biostatistics and Bioinformatics, Duke University, NC, 27703, USA; [3]Department of Electrical and Computer Engineering, Duke University, NC, 27703, USA; [4]Department of Biomedical Engineering, Duke University, NC, 27703, USA; [5]Department of Surgery, Duke University, NC, 27703, USA; [6]Department of Computer Science, Duke University, NC, 27703, USA

Corresponding author: Hanxue Gu, BS, 2424 Erwin Rd, Durham, NC 27705, Hanxue.gu@duke.edu

# Abstract

The quantity and quality of muscles are increasingly recognized as important predictors of health outcomes. While MRI offers a valuable modality for such assessments, obtaining precise quantitative measurements of musculature remains challenging. This study aimed to develop a publicly available model for muscle segmentation in MRIs and demonstrate its applicability across various anatomical locations and imaging sequences. A total of 362 MRIs from 160 patients at a single tertiary center (Duke University Health System, 2016–2020) were included, with 316 MRIs from 114 patients used for model development. The model was tested on two separate sets: one with 28 MRIs representing common sequence types, achieving an average Dice Similarity Coefficient (DSC) of 88.45%, and another with 18 MRIs featuring less frequent sequences and abnormalities such as muscular atrophy, hardware, and significant noise, achieving 86.21% DSC. These results demonstrate the feasibility of a fully automated deep learning algorithm for segmenting muscles on MRI across diverse settings. The public release of this model enables consistent, reproducible research into the relationship between musculature and health.

# Introduction

Muscle quantity and quality impact many long-term health outcomes and have been associated with metabolic function, insulin sensitivity, blood pressure, and cardiovascular disease[1,2]. However, the potential use of musculature measurements as diagnostic and prognostic biomarkers is limited by current assessment methods. Traditional non-imaging approaches, such as grip/force-based measurements, gait analysis, or bioimpedance require specialized measurements and often lack anatomical specificity[3]. Imaging-based approaches provide a more direct, quantifiable assessment of muscle structure and quality, offering greater potential for clinical and research applications[4]. Computed tomography (CT), dual X-ray energy absorptiometry (DXA), and magnetic resonance imaging (MRI) are the three most used techniques for assessing skeletal muscle size and quality in vivo[5]. While DXA is not suited for evaluating muscle quality[6], several CT-based muscle quantification algorithms have been proposed[7,8]. However, CT provides limited soft tissue contrast within muscle, restricting its ability to analyse muscle fibers and fat distribution in detail. More importantly, most CT-based methods focus on the abdominal region while ignoring other important muscle groups -- such as arm muscles for grip strength and leg muscles for mobility[8–10]. MRIs have the advantage of being widely used for the assessment of the abdomen as well as a variety of musculoskeletal conditions, resulting in a very large number of available exams spanning the entire body[11,12]. Harnessing this wealth of data could allow for broader studies of the relationship between musculature and health and eventually lead to effective body composition-based biomarkers of health.

However, the analysis of muscles in MRIs is riddled with difficulties, leading to the very limited use of MRI for this purpose. The primary issue is the non-quantitative nature of the MRIs. Muscle tissue as well as surrounding tissues can have widely varying pixel values depending on the scanner manufacturer, scanner model, sequence type, and acquisition parameters. This renders any pixel intensity-only methods (e.g., those designed for CT) highly ineffective[13]. Furthermore, muscles vary significantly in shape and size across different locations in the body and among patients. The combination of these factors results in very limited use of automated quantification and overall limited quantification of musculature in MRIs. Most existing methods are limited to single-target, region-specific segmentations, typically focusing on areas such as the thigh, and often target only a narrow set of muscle groups and MRI sequence types, which makes them difficult to generalize[14–16]. While recent models like TotalSegmentator-MRI have expanded coverage to include more muscle areas, to the best of our knowledge, they still fall short in key regions[17]. For example, they do not support critical muscle structures in the abdominal region, lower back, and leg, nor do they capture small, intricate muscles around the fingers and toes on standard clinical MRIs. Finally, TotalSegmentator-MRI's whole-skeleton muscle segmentation is designed for whole-body MRI scans, which are rarely acquired.

In this study, we propose a publicly-available MRI deep learning model that can accurately segment muscles in any body location. We thoroughly evaluate this model for typical MRI cases, rare MRI cases, and in the context of general muscle quantity assessment.

# Methods

**Study Approval**

This retrospective study was approved by the Duke University Institutional Review Board, with a waiver of informed consent due to the use of de-identified data and retrospective research.

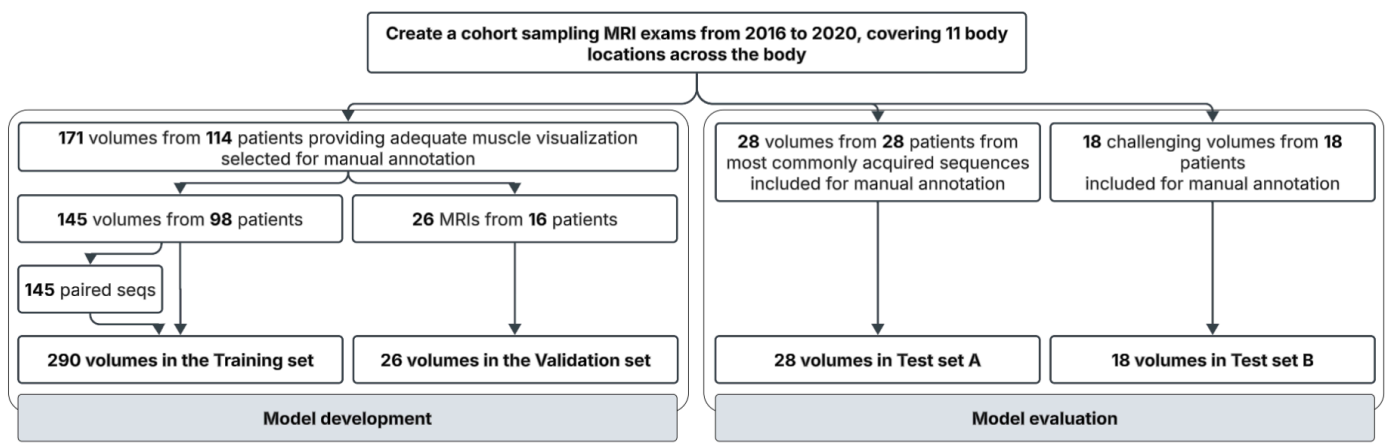

*Figure 1 Data for the model development set used for training and validation, and two independent test sets used for evaluation.*

**Datasets and Image Annotation**

In this study, we obtained a retrospective cohort consisting of a subset of patients who underwent MRI at Duke Health System between 2016 and 2020. This cohort was sampled to ensure a wide coverage of MRI imaging protocols and body locations. Specifically, we selected exams spanning 11 distinct body location categories—*Chest*, *Abdomen*, *Thoracic Spine*, *Lumbar Spine*, *Shoulder*, *Humerus*, *Hip*, *Thigh*, *Knee*, *Lower Leg*, and *Miscellaneous* (e.g., hand, wrist, foot, and ankle)-which together encompass the majority of clinically relevant skeletal muscle regions throughout the body. Exams that primarily focused on the head and face (e.g., brain and cervical spine MRIs), and those encompassing multiple areas (e.g., entire arms, legs, etc.) were excluded due to identification concerns or ambiguous region separation. From the eligible exams, we created one model development set for training and validation, and two independent test sets, the entire data inclusion/exclusion criteria are shown in Figure 1. For the development set, we selected sequences that provided adequate visualization of muscle structures (e.g., T1, T2, DIXON) to include, guided by the expertise of an experienced radiologist. Details of the sequence types included in the training/validation set are provided in eTable1. To comprehensively evaluate model performance across different data distributions, we constructed two distinct test sets. Test Set A included the more commonly-acquired sequences optimized for muscle visualization, while Test Set B comprised more challenging cases with less common sequences, muscular abnormalities, hardware artifacts, and image noise. Detailed criteria for the dataset split are provided in eMethod1, with an overview of the dataset visualization in Figure 2.

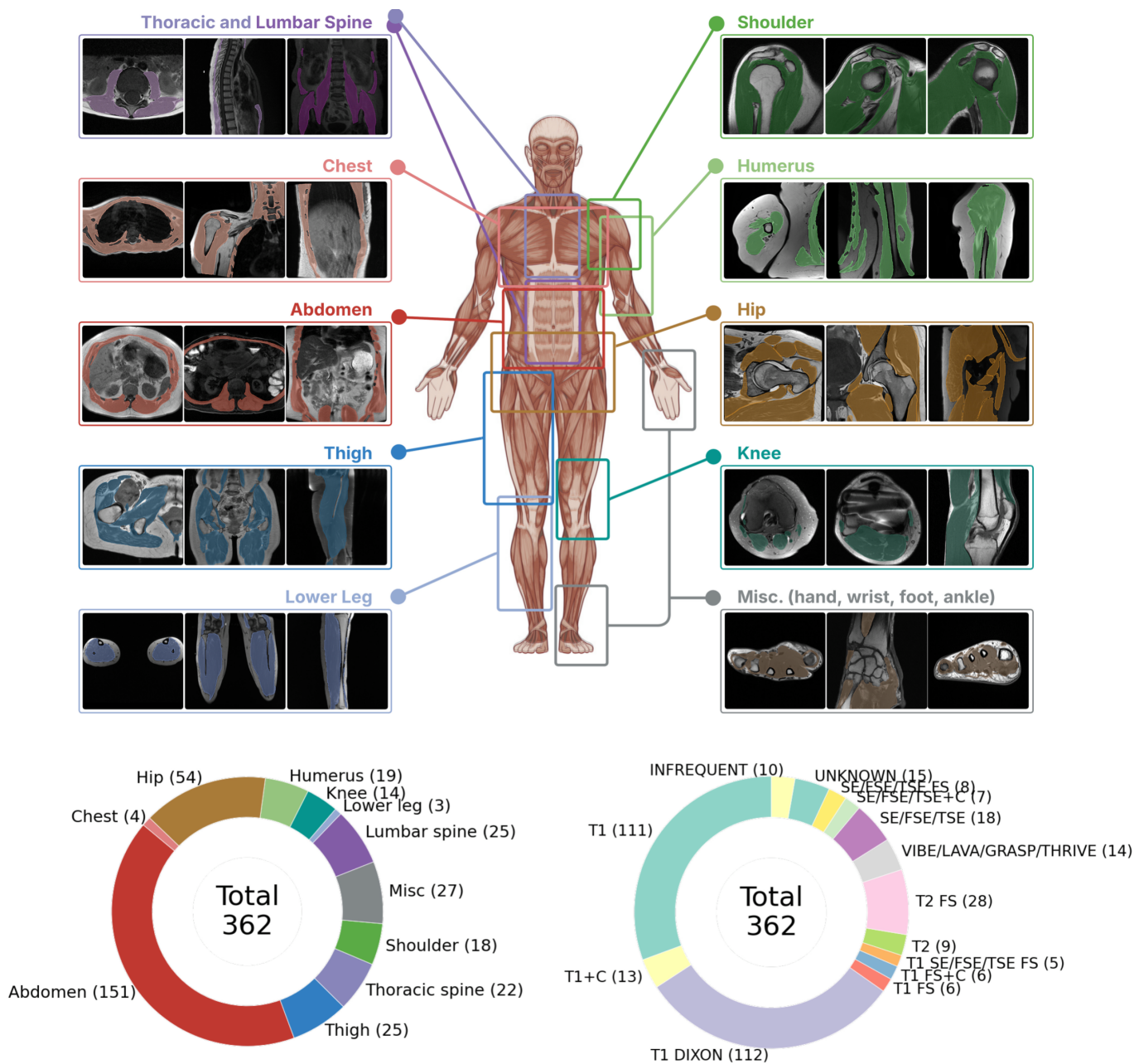

*Figure 2 The overview of full MRI dataset used for model development and evaluation. The top panel shows representative MRI slices and manual muscle annotations across 11 body regions included in the dataset; the human illustration was created by iStock.com/red_frog. The bottom left panel shows the distribution of annotated MRI exams by body location, and the bottom right panel shows the distribution by image sequence type. The summary of demographics, anatomical, and imaging characteristics across training, validation, and test sets are provided in eTable1 and more detailed composition of test sets are in eTable2 in Supplement 1.*

Annotations were performed on all selected exams by trained researchers under the guidance of experienced radiologists and subsequently reviewed and approved by one of two collaborating senior radiologists. The resulting annotations were binary—labelling each pixel of volume as either muscle or non-muscle. The annotation process details are provided in eMethod2.1 in Supplement1.

To further increase the sequence diversity of the model development set without additional annotation effort, we implemented a pairing strategy. Specifically, for each annotated exam, unannotated exams from the same patient and view were identified. Then, we included the additional images, along with the masks generated for the index images, as additional training examples following manual verification of the alignment to ensure that the annotation mask could be directly transferred to the new exams without modification. The details of this strategy are described in eMethod2.2, with a visual example shown in eFigure1 in Supplement1.

**Segmentation Model Architectures**

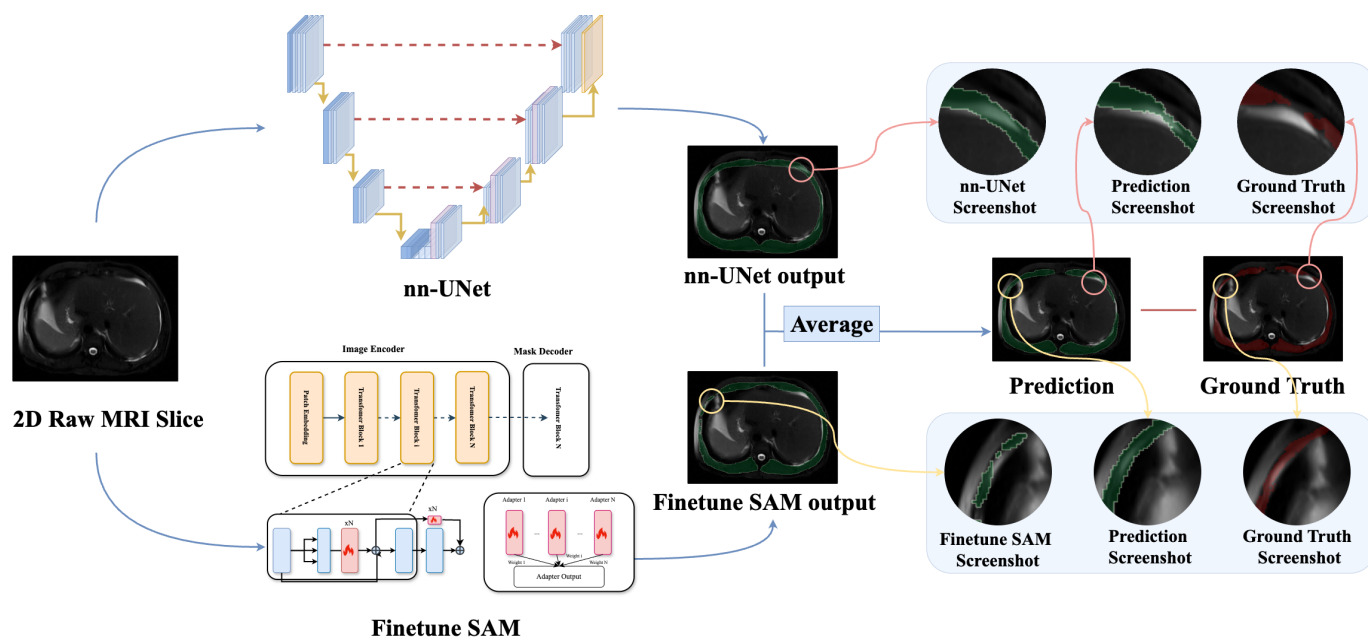

*Figure 3 Pipeline of the proposed method.*

We designed our deep learning models based on two state-of-the-art segmentation algorithms: nnU-Net and fine-tuned SAM. We also implemented an ensemble method combining the outputs of the two. nnU-Net[18] is a convolutional neural network-based algorithm that can automatically configure a matching U-Net-based segmentation pipeline based on the "fingerprint" of a dataset, including characteristics such as median shape, distribution of spacing, and intensity distribution. Fine-tuned SAM adjusts the weights of the Segment Anything Model (SAM)[19] by inserting an Adapter layer at each transformer layer[20,21]. We further proposed the *Mixture of Adapters* method, which utilizes multiple Adapter layers instead of one, and learns to differently weigh the output of each[22]. The ensemble method simply averages the probability maps generated by the nnU-Net and fine-tuned SAM, as shown in Figure 3. For comparison, we also used several strong baselines, including UNet, AttentionUNet, and SwinUNet[23–25], as described in follows:

- UNet is a convolutional neural network-based architecture that includes an encoder to compress images into high-level features and a decoder that predicts the mask based on these features. The authors also introduced the skip-connection layer that propagates features from the encoder layers to the corresponding decoder layers.
- AttentionUNet introduced Attention Gates (AG) to each decoder layer. AG computes the similarity between the current decoder's input and the encoder features from the skip connection layer. It aimed to capture only relevant information during mask prediction.
- SwinUNet replaces the encoder part of the Unet with the Swin Transformer that computes self-attention in an efficient shifted window partitioning scheme.

**Image Preprocessing**

Each MRI scan was a 3D volume in the NIFTI format. We first converted the 3D volume into a set of 2D slices. Each slice was then transformed using min-max normalization to 0-255 and resized to 1024x1024.

**Model Training and Validation**

nnU-Net was trained for 1,000 epochs with an initial learning rate of 0.01 following its default settings. Fine-tuned SAM was trained for 500 epochs with an initial learning rate of 0.0001. For each algorithm, the checkpoint achieving the highest validation score was selected for inference on the test set. Data augmentation strategies, including color jittering, random resizing and cropping, and random rotation, were utilized for both algorithms. The code was written in Python 3.10, and the code and the weights are publicly available at https://github.com/mazurowski-

lab/SegmentAnyMuscle. All models were trained using PyTorch. The training was performed on a single A6000 with 48 GB of VRAM and an AMD EPYC 7302 (16 Core) with 512 GB of RAM.

**Segmentation Accuracy Evaluation Metric**

The Dice Similarity Coefficient (DSC) was used as an evaluation metric for the muscle segmentation task. DSC reflects the degree of overlap between the area annotated as the ground truth (A) and the predicted area output by the network (B); a DSC of 1 indicates that A and B have perfect overlap, whereas a DSC of 0 indicates that there is no overlap.

In this study, True-positive (TP) and true-negative (TN) reflects the number of pixels that both A and B identify as foreground/background; false-positive (FP) reflects the number of pixels where A is negative while B predicts positive, and false-negative (FN) reflects the reverse condition. The formula for DSC is as follows:

$$DSC = \frac{2TP}{2TP + FP + FN}$$

Additionally, we report the harmonic mean of sensitivity and specificity (HSS),

$$Sensitivity = \frac{TP}{TP + FN}, Specificity = \frac{TN}{TN + FP}$$

**Estimation of Skeletal Muscle Index**

Muscle size related biomarkers have been shown to be effective predictors of various diseases, with the skeletal muscle index being one of the most widely studied features[26]. SMI is defined as Skeletal Muscle Volume (SMV) at a standard anatomical location (e.g., L3 level) divided by the patient's height, with SMV, as the key component for SMI, calculated as the volume of the muscle tissue. However, muscle distribution can vary significantly across multiple body locations. Thus, we calculate the SMV across multiple body locations to provide a more comprehensive evaluation of our model.

**Statistical Analysis**

The Wilcoxon Signed-Rank Test was used to compare the performance of the ensemble method (from nnU-Net and fine-tuned SAM) against nnU-Net and fine-tuned SAM, respectively. The association of predicted SMM with the ground truth SMM was evaluated using univariable logistic regression analyses. All statistical metrics and curves were generated using scikit-learn and SciPy package in Python. A one-sided $p < 0.05$ was considered statistically significant.

# Results

**Patient Characteristics and MRI Attributes**

The location and sequence distribution of the entire set (model development and model evaluation), as well as visual examples of annotations, are shown in Figure 2. The entire dataset included 160 patients with a median age of 54 years (range: 1-90*; patient age is capped at 90). This set comprised 88 female (55.00%) and 72 male (45.00%) patients, with most patients self-identifying as White (109, 68.13%), followed by Black (35, 21.88%), Asian (6, 3.75%), and other (4, 2.50%). The race of 6 patients (3.75%) was unknown. The dataset contained 362 MRI exams across 11 body locations, with the abdomen being the most prevalent (151 exams, 41.71%). It included 19 different MRI sequences, among which T1 Dixon, T1, and T2 fat-saturated were the most common, accounting for 112 (30.94%), 111 (30.66%), and 28 (7.73%) exams, respectively. The pairing strategy further augmented the model development set with 301 additional exams, with the most frequent sequences being T1 (66 exams, 21.93%), Dixon (57 exams, 18.94%), and Dixon in-phase (14 exams, 4.65%).Test Set A included the most common sequences for seven body parts, with T1 as the primary sequence except in the abdomen, where T1 Dixon Water (axial) and HASTE (coronal) were used. Test Set B assessed model generalizability using 11 less-common sequences, including T2 and PD, and included three muscular atrophy cases (potentially disuse or diabetes-related), two with hardware artifacts, and two with image noise. The detailed composition of each dataset is shown in eTable 1 in supplement 1.

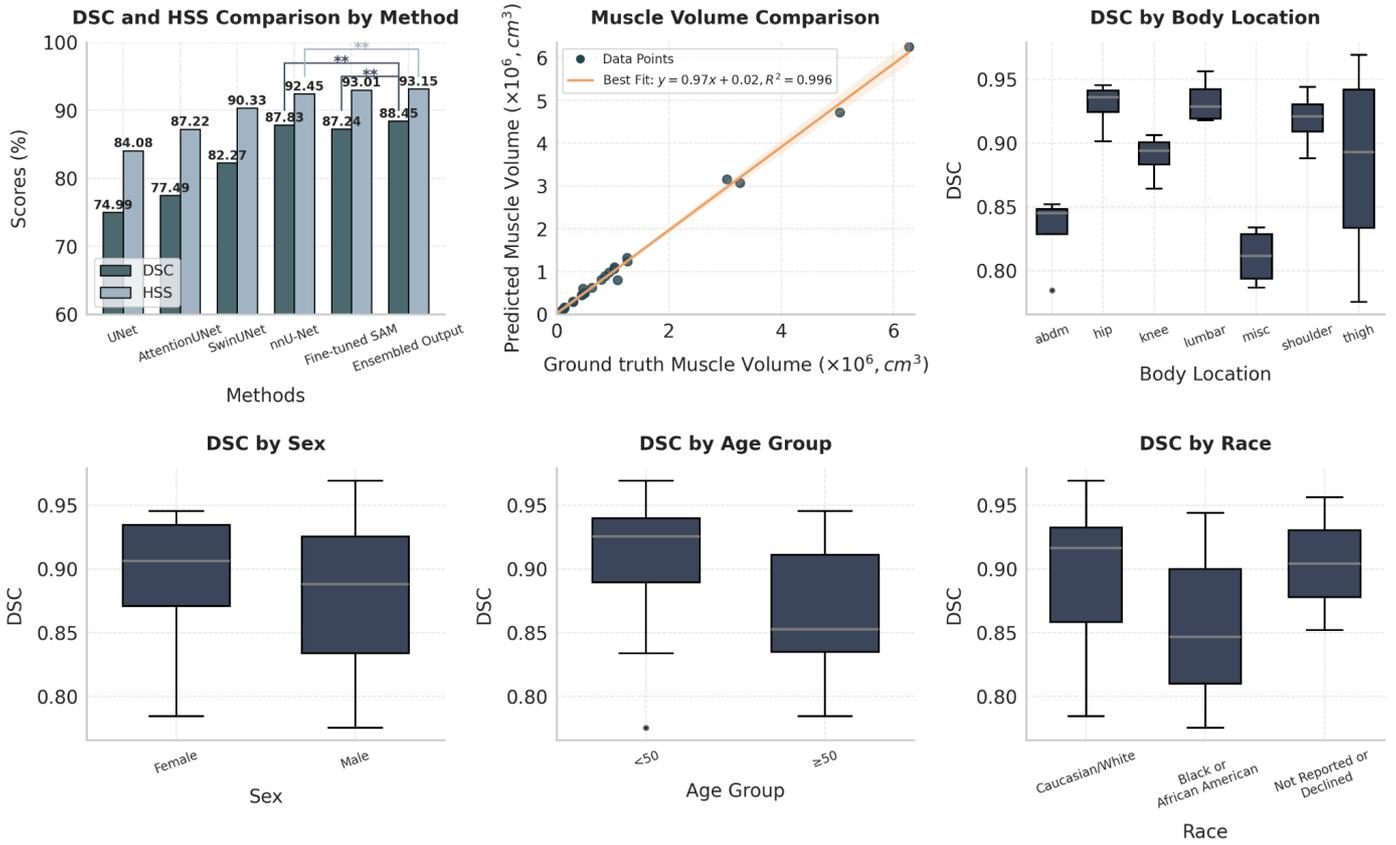

Figure 4 The performance of the models in Test Set A across subgroup analyses.

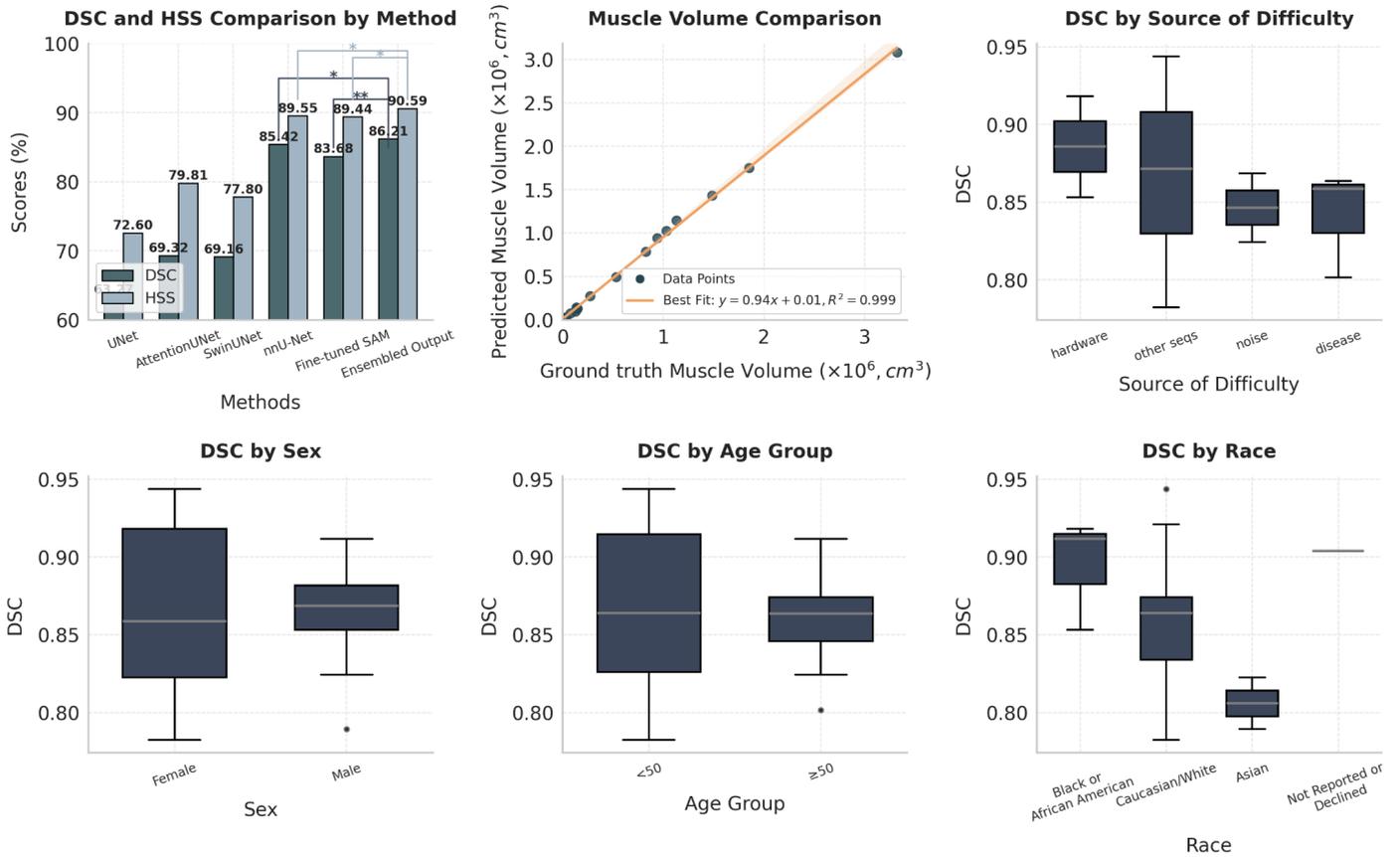

Figure 5 : The performance of the models in Test Set B across subgroup analyses.

**Segmentation Accuracy on Test Sets A and B**

The performance of nnU-Net, fine-tuned SAM, the ensemble method, and other competing methods is shown in the top-left panels of Figures 4 and 5, corresponding to Test Set A and B, respectively. nnU-Net achieved 87.83%/85.42% DSC and 92.45%/89.55% HSS, and fine-tuned SAM achieved 87.24%/83.68% DSC and 93.01%/89.44% HSS on Test Set A/B. Both models have outperformed the remaining competing methods by a large margin. The ensemble method achieved 88.45%/86.21% DSC and 93.15%/90.59% HSS. This performance was statistically significantly higher than nnU-Net and fine-tuned SAM in terms of DSC. Lastly, we demonstrate that the ensemble method is robust across body locations (for Test Set A), various challenging scenarios (for Test Set B), and demographics (including sex, age, and race). Representative examples of segmentation results are shown in Figure6.

**Skeletal Muscle Mass Measurement Comparisons**

With the high segmentation accuracy, the ensembled prediction showed high correlation with the ground truth values (Pearson $r \geq 0.99$, $P < 0.001$) for both test sets, as shown in the top-middle panels of Figure 4 and 5.

# Discussion

In this study, we successfully developed, validated, and publicly shared a deep learning-based segmentation model capable of accurately segmenting muscles on MRIs across various body locations and sequence types. This was achieved by constructing one of the largest labeled MRI muscle datasets to date, consisting of 316 MRIs from 114 patients. Our dataset spans all body regions except those covering the human face and includes a wide range of MRI sequences acquired by our proposed pairing strategy. Our algorithm achieved an impressively high DSC of 88.45% on the most frequent sequences (Test Set A) and 86.21% on more challenging cases (Test Set B).

Our model demonstrated robustness against several variables, including an exam's location in the body, patients' demographics, and different sources of difficulty. For example, as shown in the hip muscular atrophy case in Figure 5, the prediction accurately segmented muscle boundaries despite some mislabeling of intermuscular fat. Moreover, it also maintained strong performance in areas with limited training data. For example, although the development set included only two wrist exams, the model still achieved a DSC of 78.66% for the wrist. Further investigation, as shown in the misc. In the case shown in Figure 5, it was revealed that while muscle structures were well-segmented, tendons in the carpal tunnel were occasionally missed.

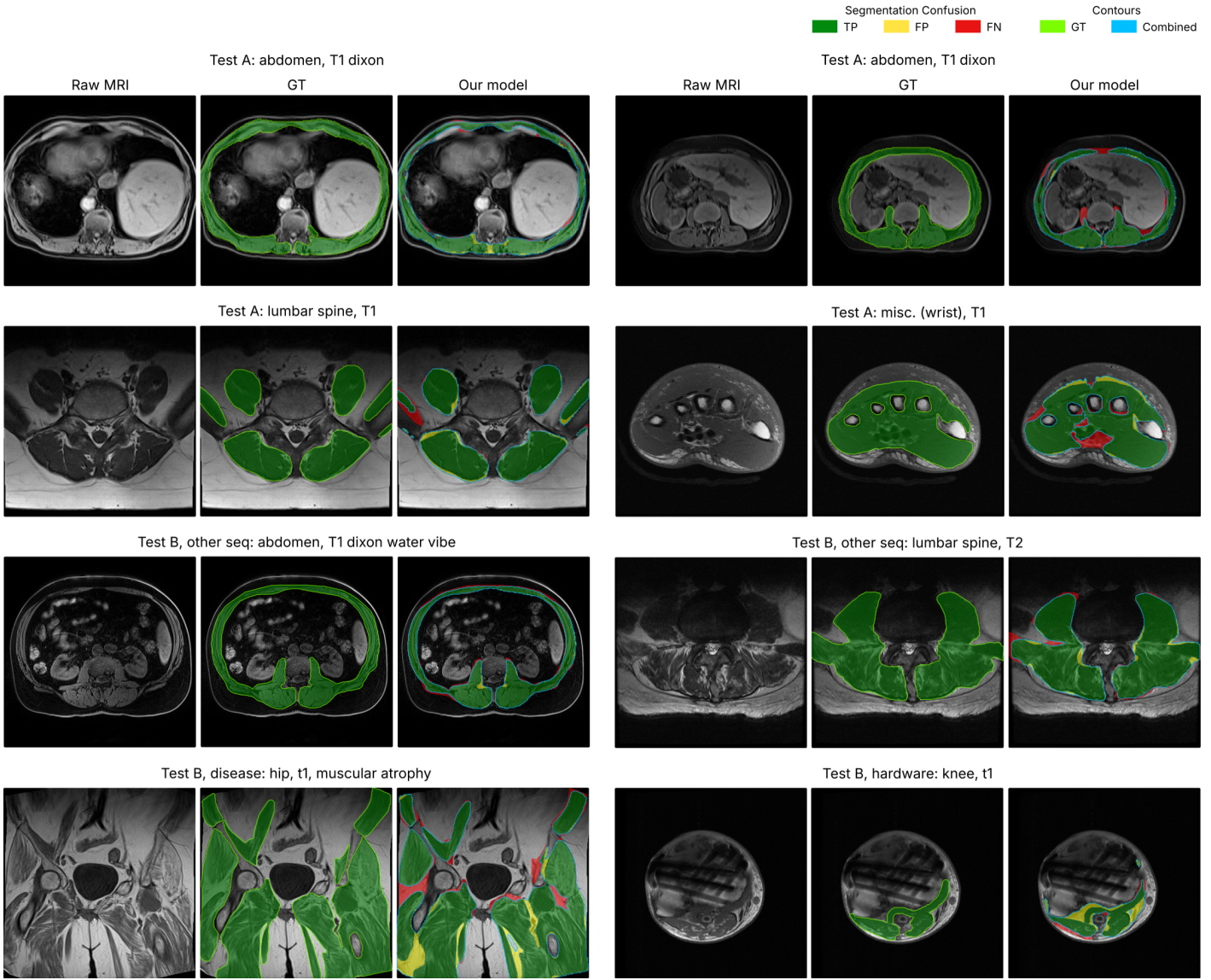

*Figure 6  The visualization of model performance in test sets.*

The success of our model not only depends on the richness of the collected dataset, but also on the proposed segmentation strategy. When compared to U-Net, a widely used approach in medical imaging applications, our model surpasses it by 13.46% and 22.94% in DSC on test set A and B, respectively. These substantial improvements highlighted the value of employing a more advanced algorithm. Our model also achieved statistically significant improvements over nnU-Net, which is directly adopted by many concurrent works, such as TotalSegmentator-MRI, and provides broader anatomical coverage, including complex and smaller muscle groups such as ankle and wrist, as well as wider MRI sequence coverage.

Beyond superior performance in accuracy, our model also offers significant clinical potential by enabling automated muscle segmentation in MRI. Manual muscle segmentation is rarely performed in routine clinical practice due to its time-consuming nature. Specifically, it took an experienced annotator 5-11 hours to annotate a full volume. By providing fast and accurate muscle segmentation across diverse body locations and sequence types, our model facilitates the downstream assessment of muscle volume and quality, unlocking new possibilities for muscle-related clinical applications. For example, our model can help identify muscle loss earlier, track muscle disease progression such as sarcopenia, and support decision-making in neuromuscular disorders and post-surgical recovery. Additionally, integrating automated muscle quantification into routine MRI analysis can facilitate the opportunistic screening of muscle health and risk factors without requiring dedicated scans, and large-scale muscle health studies over large populations, uncovering trends in muscle aging, and disease progression. Such insights could contribute to improving risk stratification methods, and health care strategies.

We shared the code and training weights publicly at https://github.com/mazurowski-lab/SegmentAnyMuscle to accelerate research in this direction. While this work does not provide direct clinical applications, the high accuracy of our segmentation model for accurate estimation of the skeletal muscle mass and the public availability of the model enable opportunities for advanced research in muscle quantification, disease monitoring, and population studies.

**Limitations**

This study has certain limitations. First, although the proposed segmentation model demonstrates high accuracy on the test sets from the same institution, tests using data from an external cohort were not performed. This suggests the need for multi-institutional data to enhance the algorithm's generalizability, or additional training using our publicly available code. Second, our method is currently limited to binary muscle segmentation. Future annotation work that converts binary masks into multi-class ones. Multi-class muscle segmentation can enable more advanced muscle assessments, such as evaluating asymmetry between left and right muscles for patients' balance analysis or linking degeneration in specific muscle groups to functional outcomes like reduced grip strength. Third, the performance of patients with diseases is sub-optimal. Lastly, only a proof-of-concept evaluation on the clinical application is conducted.

# Conclusions

We developed a segmentation model that can segment muscles from MRIs. In this retrospective study, the segmentation accuracy of the model was high and was maintained when the body location and sequence type of MRI as well as patient demographics changed. Muscle segmentation for MRI is an important fundamental technology for various research areas, including body composition interpretation, development areas in surgical planning, and risk stratification. The proposed model can be applied to any related research smoothly given its public availability. It can not only measure muscle quantitatively but also be used in conjunction with other computer-assisted methods to aid in annotation.

# Acknowledgments

*Competing interest declaration:* no conflict of interest

*Code availability: code is shared publicly available on:* https://github.com/mazurowski-lab/SegmentAnyMuscle.

*Data availability: data may be shared publicly available upon publication and after getting the institutional consent.*

21. Gu H, Dong H, Yang J, Mazurowski MA. How to build the best medical image segmentation algorithm using foundation models: a comprehensive empirical study with Segment Anything Model. Published online May 13, 2024. doi:10.48550/arXiv.2404.09957

22. Wang Y, Agarwal S, Mukherjee S, et al. AdaMix: Mixture-of-Adaptations for Parameter-efficient Model Tuning. Published online 2022. doi:10.48550/ARXIV.2205.12410

23. Chen J, Lu Y, Yu Q, et al. TransUNet: Transformers Make Strong Encoders for Medical Image Segmentation. Published online February 8, 2021. doi:10.48550/arXiv.2102.04306

24. Ronneberger O, Fischer P, Brox T. U-Net: Convolutional Networks for Biomedical Image Segmentation. *Lect Notes Comput Sci Subser Lect Notes Artif Intell Lect Notes Bioinforma*. 2015;9351:234-241.

25. Oktay O, Schlemper J, Folgoc LL, et al. Attention U-Net: Learning Where to Look for the Pancreas. Published online 2018. doi:10.48550/ARXIV.1804.03999

26. Skeletal Muscle Mass Index - an overview | ScienceDirect Topics. Accessed May 12, 2025. https://www.sciencedirect.com/topics/medicine-and-dentistry/skeletal-muscle-mass-index
16

# Supplement

**eMethod1: Dataset Preprocessing Process**

1. Data Collection and De-identification

The dataset includes MRI exams from patients at Duke Health System collected between 2016 and 2020. All exams were de-identified using the honest broker service within Duke's secure, cloud-based Protected Analytics Computing Environment (PACE). Exams that could not be processed through this de-identification pipeline were excluded.

2. Body Location Category Extraction

MRIs were categorized into 11 distinct body parts (e.g., *Chest*, *Abdomen*, *Thoracic Spine*, *Lumbar Spine*, *Shoulder*, *Humerus*, *Hip*, *Thigh*, *Knee*, *Lower Leg*, and *Miscellaneous Parts* (e.g., hand, wrist, foot, and ankle)) using location-related keywords from the protocol descriptions. For example, an exam with the protocol description "MRI hip left with and without contrast" was assigned to the "Hip" category. Note that MRIs covering head and face areas (e.g., brain and cervical spine) were excluded to prevent potential protected health information leaks. Also, MRIs composing multiple areas (e.g., entire arms, legs, spines, and body) were excluded for analysis purposes.

3. Sequence Category and Frequency Extraction

MRI series categories were extracted using the SeriesDescription field from the DICOM header, which contains key information about the MRI sequence, fat saturation status, phase, and contrast use. Each component was identified mainly through keyword detection within the description.

The classification of MRI sequences included the following terms: "t1" (T1 Weighted), "t2" (T2 Weighted), and "pd" (Proton Density) sequences, along with specialized techniques such as "dixon" (fat/water separation), "stir" (Short Tau Inversion Recovery), "tirm" (Turbo Inversion Recovery Magnitude), "vibe" (Volumetric Interpolated Breath-hold Examination), "lava" (Liver Acquisition with Volume Acceleration), "grasp" (Golden-Angle Radial Sparse Parallel), "thrive", "se" (Spin Echo), "fse" (Fast Spin Echo), "tse" (Turbo Spin Echo), and "mra" (Magnetic Resonance Angiography). Fat saturation status was classified as "fs" for fat-saturated and "no fs" for non-fat-saturated sequences. Phase information distinguished water phase and in-phase/out-phase images, identified by the keywords "WATER", "in phase" and "out phase", respectively, while for dixon, four distinct sequences (in-phase, out-of-phase, water, and fat) were identified. Contrast use, indicating post-contrast imaging, was determined by the presence of "c" in the description. The MRI sequence classification was further simplified for reporting purposes by grouping similar techniques under a unified category, as different manufacturers use varying terminology. For instance, motion-reduction techniques such as "vibe/lava/grasp/thrive" were grouped together, as were "se/fse/tse" sequences for spin-echo variants and "c" for post-contrast variants such as dynamic, equilibrium, portal venous, and breath-hold.

MRI sequences that do not display muscles under adequate visualization quality, such as diffusion, localizer, and mpr/mrcp/twist reconstruction, were excluded in our study due to the poor quality and the lack of clinical relevance for muscle analysis.

The prevalence of MRI sequences across different body regions was analyzed. In Test Set A, the most commonly acquired sequences with optimal muscle visualization are identified. T1 was the predominant sequence for all body locations except for the abdomen. For abdomen, sequence usage varied significantly across views; T1 Dixon Water (axial) and HASTE (coronal) were the most frequently used. Additionally, in Test Set B, 11 less commonly acquired sequences were identified by selecting the second- through fourth-most frequently used sequences for each body part. One volume was chosen per sequence, prioritizing the most frequently acquired view and the body region with the largest patient distribution when sequences appeared multiple times.

4. Disease, Hardware, and Image Quality Assessment in Test Set B

Test Set B included a diverse range of challenging cases that could compromise both diagnostic interpretation and segmentation performance, particularly for muscle assessment. These cases featured MRI exams with muscle-affecting diseases, hardware artifacts, and image noise, in addition to less frequently acquired sequences compared to T1.

All cases were manually identified by experienced radiologists. For disease cases, patients with conditions affecting muscle quantity, quality, or morphology were first selected based on radiology reports. From this subset, radiologists further identified cases with the most pronounced muscular abnormalities. For hardware artifacts, cases with significant imaging disturbances due to implants were similarly chosen through expert review. For image noise, MRI scans with moderate-to-poor image quality due to signal-to-noise (SNR) issues were included, as these artifacts limited muscle visibility and posed challenges for accurate annotation.

**eMethod2: Dataset Annotation Process**

1. Manual Annotation

Muscle structures were defined to include all visible skeletal muscle tissue, including tendons, to ensure consistent representation across body regions. For instance, if an abdominal exam included parts of the arms, all visible muscles were annotated to support the development of a robust segmentation model capable of recognizing muscles regardless of location.

A binary annotation approach was implemented, classifying areas as either "muscle" or "non-muscle" to develop a binary segmentation model, with future plans for multi-class segmentation. A pixel was annotated as muscle if radiologists determined it had more than a 50% probability of belonging to a muscle. However, certain cases, such as those in Test Set B, presented challenges due to ambiguities in defining muscle regions. Examples include handling artifacts that obscure muscle visibility and determining whether intramuscular fat in diseased, highly fatty muscles should be annotated. These cases involved subjective judgment, and annotations were standardized based on radiologists' preferences to maintain consistency.

Manual annotation was performed by a multi-tier team under the supervision of two fellowship-trained musculoskeletal radiologists to ensure accuracy and consistency. The annotation process followed a structured workflow: (1) Initial annotation by annotators with limited radiology experience, requiring approximately 4–8 hours per volume; (2) Cross-review within the annotation team to check quality for all exams; (3) Review by experienced readers covering all cases; and (4) Refinement and final approval, adding an additional 1–3 hours per volume. In total, the manual annotation process took approximately 5–11 hours per volume.

Due to the substantial workload and limited availability of experienced annotators, we established selection criteria for MRI volumes for the model development dataset. Initial efforts prioritized MRI exams from 2018, focusing first on Abdomen and Hip. Annotations were later expanded to other body regions with fewer cases per location, including additional abdomen exams from other years. Ultimately, annotated data covered 11 body locations, enhancing the diversity and applicability of the dataset. Similarly, T1 sequences were prioritized, but not limited, for their widespread use in MRI exams and superior clarity for muscle segmentation.

2. Annotation Transfer and Pairing

To enhance dataset diversity without introducing additional manual annotations, we transferred existing annotations from sequences to other aligned sequences within the same patient and view. A research annotator manually verified alignment to ensure that the annotation mask could be directly applied without modification. Only well-aligned pairs were included in the training set, with examples illustrated in eFigure 1.

This expansion, referred to as paired data, added 145 MRI exams to the training set, incorporating additional sequences such as T2, TIRM, and PD-weighted images, increasing the training set from 145 T1-prioritized volumes to 290 volumes across various sequences.

**eFigure1: The pipeline for extending the dataset to non-T1 sequences**

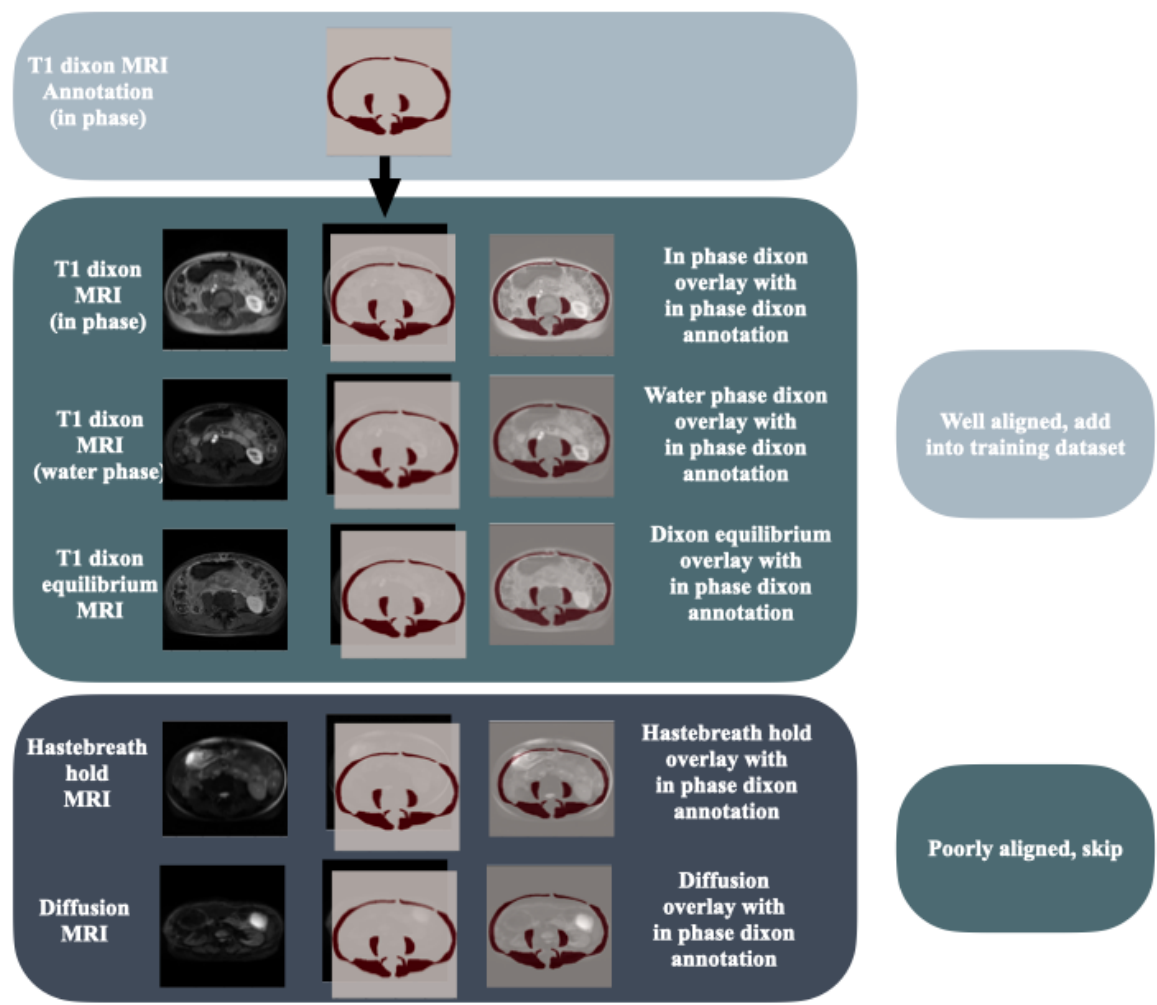

eTable1: The summary of demographic, anatomical, and imaging characteristics of training, validation, and test sets

|  | Training: vol (pt) | Validation: vol (pt) | Test A: vol (pt) | Test B: vol (pt) |
|---|---|---|---|---|
| Total | 290 (98) | 26 (16) | 28 (28) | 18 (18) |

| Demographic | Training | Validation | Test A | Test B |
|---|---|---|---|---|
| Age: mean | 50.21+-22.66 | 54.5+-21.90 | 50.07+-21.83 | 48.44+-19.68 |
| Age: range | (1, 90*) | (14, 86) | (9, 86) | (23, 88) |
| Sex (Female): n (%) | 58 (59.18) | 10(62.50) | 11 (39.29) | 9 (50.0) |
| Sex (Male): n (%) | 40 (40.82) | 6 (37.50) | 17 (60.71) | 9 (50.0) |
| Sex (N/A): n (%) | - | - | - | - |
| Race (White): n(%) | 65 (66.33) | 13 (81.25) | 19 (67.86) | 12 (66.67) |
| Race (Black): n (%) | 22 (22.45) | 3 (18.75) | 7 (25.00) | 3 (16.67) |
| Race (Asian): n (%) | 4 (4.08) | - | - | 2 (11.11) |
| Race (Other): n (%) | 3 (3.06) | - | - | - |
| Race (N/A): n (%) | 4 (4.08) | - | 2 (7.14) | 1 (5.55) |

| Body Location | Training: vol (pt) | Validation: vol (pt) | Test A: vol (pt) | Test B: vol (pt) |
|---|---|---|---|---|
| abdomen | 134 (28) | 11 (6) | 4 (4) | 2 (2) |
| chest | 4 (3) | - | - | - |
| hip | 40 (12) | 6 (3) | 4 (4) | 4 (4) |
| shoulder | 10 (6) | 2 (2) | 4 (4) | 2 (2) |
| humerus | 19 (7) | - | - | - |
| thigh | 18 (4) | 2 (1) | 4 (4) | 1 (1) |
| knee | 7 (6) | 1 (1) | 4 (4) | 2 (2) |
| lower leg | 3 (1) | - | - | - |
| thoracic spine | 21 (11) | 1 (1) | - | - |
| lumbar spine | 16 (10) | 2 (1) | 4 (4) | 3 (3) |
| misc | 18 (10) | 1 (1) | 4 (4) | 4 (4) |

| MRI Sequence | Training: vol (pt) | Validation: vol (pt) | Test A: vol (pt) | Test B: vol (pt) |
|---|---|---|---|---|
| t1 | 66 (28) | 14 (10) | 24 (24) | 7 (7) |
| t1+c | 12 (10) | - | - | 1 (1) |
| t1 dixon | 103 (25) | 5 (5) | 2 (2) | 2 (2) |
| t1 fs | 5 (4) | - | - | 1 (1) |
| t1 fs+c | 6 (5) | - | - | - |
| t1 se/fse/tse fs | 5 (3) | - | - | - |
| t2 | 8 (7) | - | - | 1 (1) |
| t2 fs | 26 (20) | 1 (1) | - | 1 (1) |
| t2 fs+c | 1 (1) | - | - | - |
| pd | 3 (2) | - | - | 1 (1) |
| pd fs | 1 (1) | - | - | 1 (1) |
| stir | | | | 1 (1) |
| vibe/lava/grasp/thrive | 12 (5) | 1 (1) | - | 1 (1) |
| se/fse/tse | 11 (10) | 5 (5) | 2 (2) | - |
| se/fse/tse+c | 6 (6) | - | - | 1 (1) |
| se/fse/tse fs | 8 (6) | - | - | - |
| se/fse/tse fs+c | 1 (1) | - | - | - |
| mra | 1 (1) | - | - | - |
| unknown | 15 (8) | - | - | - |

**eTable2: The detailed composition of Test Set A (frequently acquired sequences by body region and MRI view) and Test Set B (challenging cases by mri sequence, body region, and artifact type)**

| | **TEST A** | | |
|---|---|---|---|
| **Body Part** | **MRI sequence (ax: cor: sag)** | **# of vols** | **Total** |
| abdomen | t1 dixon (t1 dixon water): se/fse/tse (haste): - | 2:2:0 | 4 |
| hip | t1: t1: - | 1:3:0 | 4 |
| shoulder | t1: t1: t1 | 1:1:2 | 4 |
| thigh | t1: t1: t1 | 1:2:1 | 4 |
| knee | t1: t1: t1 | 1:1:2 | 4 |
| lumbar spine | t1: -: t1 | 2:0:2 | 4 |
| misc | t1: -: - | 4:0:0 | 4 |

**TEST B**

| Challenges | MRI sequence | Body Part | Total |
|---|---|---|---|
| various seqs | t1+c | misc | 1 |
| | t1 dixon (t1 dixon in phase) | thigh | 1 |
| | t1 dixon (t1 dixon water vibe) | abdomen | 1 |
| | t1 fs | shoulder | 1 |
| | t2 | lumbar spine | 1 |
| | t2 fs | misc | 1 |
| | stir | lumbar spine | 1 |
| | pd | knee | 1 |
| | pd fs | shoulder | 1 |
| | vibe | hip | 1 |
| | se/fse/tse+c | abdomen | 1 |
| disease | - | hip, foot | 3 |
| hardware | - | knee, lumbar | 2 |
| image noises | - | hip | 2 |